\documentclass[aps,prl,twocolumn,showpacs,superscriptaddress]{revtex4-1}

\usepackage{graphicx}
\usepackage{times}
\usepackage{bbm}
\usepackage{color}
\usepackage{units}
\usepackage{subfigure}

\usepackage{amsmath, amsthm, amssymb}

\newcommand{\eref}[1]{Eq.~(\ref{#1})} 
\newcommand{\fref}[1]{Fig.~\ref{#1}}  

\newcommand{\eig}{\ensuremath{\mathrm{eig}}}
\newcommand{\rmi}{\mathrm{i}}

\newcommand{\comment}[1]{}

\begin{document}
\title{Preparing the bound instance of quantum entanglement}		 

\author{J.~DiGuglielmo}
\author{A.~Samblowski}
\author{B.~Hage}
\affiliation{Institut f\"{u}r Gravitationsphysik, Leibniz Universit\"{a}t Hannover, 30167 
Hannover, Germany and \\ Max-Planck Institut f\"{u}r Gravitationsphysik,
30167 Hannover, Germany}

\author{C.~Pineda}
\affiliation{Instituto de F\'{\i}sica, Universidad Nacional Aut\'onoma de M\'exico, M\'exico}
\affiliation{Institute of Physics and Astronomy, University of Potsdam, 14476 Potsdam, Germany}

\author{J.~Eisert}
\affiliation{Dahlem Center for Complex Quantum Systems, 
Freie Universit{\"a}t Berlin, 14195 Berlin, Germany}
\affiliation{Institute of Physics and Astronomy, University of Potsdam, 14476 Potsdam, Germany}

\author{R.~Schnabel}
 \affiliation{Institut f\"{u}r Gravitationsphysik, Leibniz Universit\"{a}t Hannover, 30167 
Hannover, Germany and \\ Max-Planck Institut f\"{u}r Gravitationsphysik,
30167 Hannover, Germany}

\date{\today}

\begin{abstract}
Among the possibly most intriguing aspects of quantum entanglement is that it comes in ``free'' and ``bound'' instances.  Bound entangled states require entangled states in preparation but, once realized, no free entanglement and therefore no pure maximally entangled pairs can locally be regained. 
Their existence hence certifies an intrinsic irreversibility of entanglement in nature and suggests a connection with thermodynamics. 
In this work, we present a first experimental unconditional preparation and detection of a bound entangled state of light. We consider continuous-variable entanglement, use convex optimization to identify regimes rendering its bound character well certifiable, and realize an experiment that continuously produced a {distributed} bound entangled state with an extraordinary and unprecedented significance of more than ten standard deviations away from both separability and distillability. Our results show that the approach chosen allows for the efficient and precise preparation of multi-mode entangled states
of light with various applications in quantum information, quantum state engineering and high precision metrology.
\end{abstract}

%
%
%
%
%
%
%

 \maketitle

\maketitle
The preparation of complex multi-mode entangled states of light {distributed to two or more parties} is a necessary starting point for applications in quantum information processing 
\cite{Distillation}
\cite{OpticalTechnologies}
as well as for fundamental physics research.  
An aggressively pursued example of the latter is the
preparation of the bound instance of entanglement, {a type of entanglement
that can only exist in higher-dimensional or 
multi-mode quantum states} \cite{Bound}.  Bound entanglement is fundamentally interesting
since, in contrast to ``free'' entanglement, it can not be distilled to form
fewer copies of more strongly entangled {pure} states \cite{Bound} by any local  device allowed by the rules of quantum mechanics. This irreversible
character has triggered entire theoretical research programmes \cite{RMP}, in
particular by linking entanglement theory to a {thermodynamical picture}, with this
irreversibility reminiscent of---but being inequivalent with---the second law of
thermodynamics 
\cite{Thermo1}.
In order to investigate such connections
both new theoretical as well as experimental means of constructing multi-mode
states must be innovated.

In recent years, great progress in information processing, metrology
and fundamental research has actually been achieved in the photon counting (discrete variable, DV) regime using postselection  
\cite{Distillation}.
States of light are the optimal systems for entanglement distribution because they propagate fast and can preserve their coherence over long distances. 
\emph{Postselection} means that the measurement outcome of the detectors which characterizes the quantum state is also used to select the state, conditioned on certain measurement outcomes. In such an approach, conditional applications are possible, however, an \emph{unconditional} application of the states in downstream experiments is conceptually not possible.
Another limitation that any postselected architecture will eventually face is that without challenging prescriptions of measurement, quantum memories and conditional feedforward, the preparation (post-selection) efficiency will exponentially decay with an increasing number of modes.
In parallel to postselected architectures of light, unconditional {platforms} for research in quantum information {have} been developed which 
{build} on the detection of position and momentum like variables having a continuous spectrum and a Gaussian statistics. 
In {such platforms} the preparation efficiency of one mode is identical to the preparation efficiency of $N$ modes. In the past, this continuous variable (CV) platform {has been used} to demonstrate the Einstein-Podolsky-Rosen (EPR) paradox 
\cite{EPR35}
and unconditional quantum teleportation \cite{tele1}.
Recently, the CV platform {has been extended} to investigate multimode entangled states 
\cite{Leuchs};
however,  the significance of their nonclassical properties have typically been smaller compared to their postselected counterparts.

In this work, we demonstrate the continuous unconditional preparation of one of the rarest types of multi-mode entangled states -- bipartite bound entangled states -- using the CV platform. The property of \emph{bound entanglement} is verified by four downstream balanced homodyne detectors with a detection efficiency of almost unity. Alternatively, our setup can make available bound entangled states for any downstream application. The bound entanglement is generated with unprecedented significance, i.e., with state preparation error bars small with respect to the distance to the free entanglement regime and with respect to the distance to the separability regime. Our result is achieved by the convex optimization of state preparation parameters, and by introducing the experimental techniques of single-sideband quantum state control and classical generation of hot squeezed states.

The first ever generation of bound entangled states was claimed in 2009  \cite{Bourennane}. This work used photon counting and postselection, however, the data presented did not support this claim, an issue which has been addressed in a comment, see Ref.\ \cite{Comment}. In {Ref.\ }\cite{Bruss} a DV nuclear magnetic resonance state whose {density} matrix has a small contribution of bound entanglement {has been observed}. Such a state {has been called} a ``pseudo-bound entangled state''. Very recently, the actual first bound entangled states have been generated in two experiments, both on the basis of discrete variables. In {Ref.\ }\cite{B1} bipartite bound entangled states of trapped ions {have been} verified by the unconditional detection of resonance fluorescence. In {Ref.\ }\cite{B2} the first bound entangled states of light have been generated, albeit of multipartite and not
of bipartite nature. Similar to Ref.\ \cite{Bourennane}, photon counting and postselection {have been used}. An unconditional application of the distributed entanglement in a downstream experiment is hence not possible. This is now made possible in our work, with a significance of bound entanglement that has not been achieved using postselection.

\begin{figure}
\includegraphics[width=7.5cm]{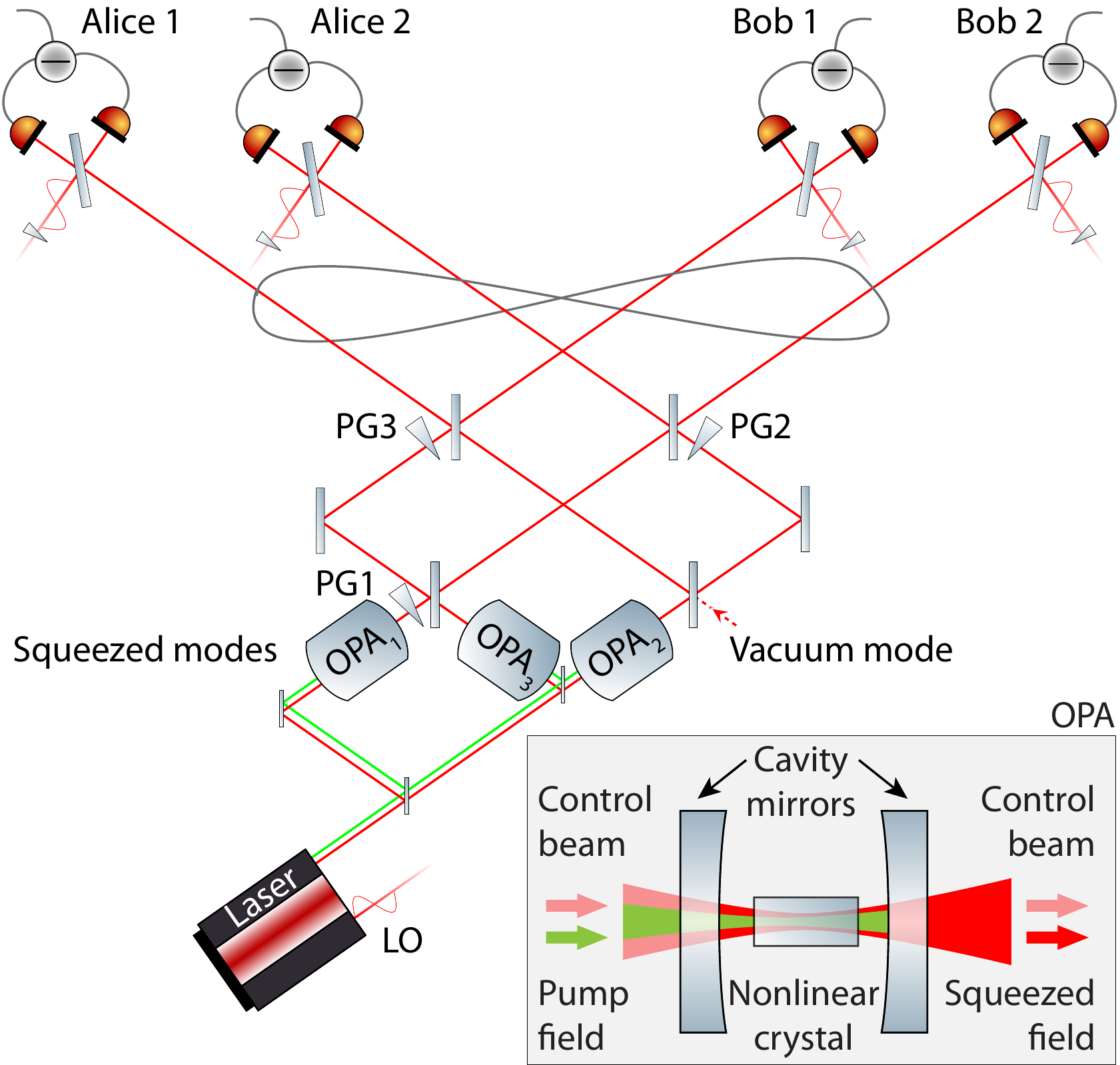}
\caption{Experimental setup: {The experiment is composed of three optical
parametric amplifiers ($\mathrm{OPA}_{1-3}$), three actively controlled piezo
mounted mirrors forming phase-gates ($\mathrm{PG1-3}$) and four homodyne
detectors which are independent of the preparation. The inset shows the
construction of an OPA as a non-linear crystal inside a resonator producing a
spatial $\mathrm{TEM}_{00}$ (transverse electro-magnetic) mode. 
The bound entangled state is obtained through the bipartite splitting such that Alice and Bob each possess
two of the four modes.}}
\label{exp_setup}
\end{figure}

Our theoretical search for CV Gaussian bound entangled states of light begins with three
(non-pure) squeezed input modes and a vacuum mode overlapped on four {beam splitters acting as} phase-gates.  This 
yields several independent parameters to be chosen that includes three pairs of
quadrature variances and the splitting ratios and the relative phases of the
phase-gates.  Additional vacuum contributions due to optical losses at different
locations in the experiment have to be considered as well. 
As it turns out, bound entanglement is extremely rare in this multi-dimensional parameter space. 
Hence, to theoretically identify suitable regimes for experimental
certification is a challenging task: Known examples of CV bound entangled states, including
those of Ref.\ \cite{werner2001}, will have both free entangled and separable
states very nearby. Optimal entanglement witnesses can be efficiently constructed
for Gaussian states \cite{P}, yet to maximize the distance of an optimal hyperplane
separating separable states to the boundary of {non-distillable} states---hence maximizing
robustness of a preparation---is a non-convex
difficult problem. What is more, a reasonable compromise with the preparation complexity
has to be found, with a surprisingly simple feasible scheme being shown in \fref{exp_setup}.

We now present the measures required for verifying the presence of bound entanglement.
Since the studied states are Gaussian they are fully described by their first---which will not play a role here---and second
moments, specified by the covariance matrix of a state $\hat \rho$ \cite{simon1987,Survey,Survey2}.  
We define a
set of quadratures for each optical mode given by $\hat x_j=(\hat a_j+\hat
a_j^\dag)/2^{1/2}$ and $\hat p_j=-i(\hat a_j-\hat a_j^\dag)/2^{1/2}$ where $\hat
a_j, \hat a_j^{\dagger}$ are the annihilation and creation operators,
respectively. Collecting these $2n$ coordinates in a vector $\hat O=(\hat x_1,\hat
p_1,\ldots,\hat x_n,\hat p_n)$,		  we can write the commutation relations as
$[\hat O_j,\hat O_k]=i\sigma_{j,k}$, where $\hbar=1$ and is a matrix
$\sigma$ often known as {\it symplectic matrix}.  The second
moments are embodied in the $2n\times 2n$ covariance matrix 
\begin{equation}
		 \gamma_{j,k}=2{\rm Re}\,{\rm tr}\left(\hat \rho(\hat O_j-d_j)(\hat O_k-d_k)\right),
\end{equation}
with $d_j = {\rm tr}(\rho \hat O_j)$, giving rise to a {real-valued} symmetric matrix
$\gamma$, see supplementary material.

Verification of  {bipartite} bound entanglement requires showing that the state is entangled (inseparable) with respect to a bipartition of the modes and that the state remains positive under partial transposition \cite{Bound,werner2001,giedkeadded} 
{proving that the state is not distillable}.  
The state is said to be {\it entangled} if no 
physical covariance matrices $\gamma_A$ and
$\gamma_B$ exist of states in modes $A$ and $B$, respectively,  so real
matrices satisfying $\gamma_A, \gamma_B  \geq -\rmi \sigma$, such that \cite{werner2001}
		 $\gamma  \geq \gamma_A \oplus \gamma_B$.
This idea suggests a natural entanglement measure \cite{EMeasure} for Gaussian states,
defined as the solution of 
\begin{eqnarray}
E(\gamma) &=&  1- \max_{\gamma_A, \gamma_B}		  x \\
\gamma & \geq &\gamma_A \oplus \gamma_B,  \quad
\gamma_A, \gamma_B \geq -\rmi x \sigma.\nonumber
\label{eq:quantify_entanglement}
\end{eqnarray}
$E(\gamma)>0$ indeed implies that the state is entangled.  The above problem is
known as a semi-definite program, a convex optimization problem
that can efficiently be solved.

{\it Non-distillability} can be tested by evaluating the partial transposition of a state 
\cite{Bound} which physically
reflects time reversal. For covariance matrices, partial transposition amounts to changing the sign of momentum
coordinates or by applying the operation  $\gamma^\Gamma= M \gamma M$, where $M=(1,1,1,1, 1,-1,1,-1)$, with a
$-1$ in all momentum coordinates belonging to $B$.  A covariance matrix
$\gamma$ is said to be PPT if its partial transpose is again a legitimate
covariance matrix, or equivalently, $\gamma^{\Gamma} + \rmi \sigma \geq 0$.  
A measure as to the quantitative extent a state is PPT can be taken to be
the minimum eigenvalue of this matrix,
\begin{equation}
P(\gamma) = \min \eig ( \gamma^{\Gamma} + \rmi \sigma ).
\label{eq:P_def}
\end{equation}
The continuity of the eigenvalues with respect to variations in the matrix are
enough to guarantee that the measure is meaningful.  A strictly positive value
of $P(\gamma)$ unambiguously certifies that the state is not distillable.

Based on our theoretical parameter search our final experimental setup is
realized as shown in \fref{exp_setup}. In total three optical parameter amplifiers (OPAs),
three phase-gates, consisting of a beam splitter and a piezo mounted mirror, and
a vacuum mode are utilized as the base setup.  The four homodyne detectors are
only necessary for the verification of bound entanglement but not for its
preparation. We set our OPAs to produce the minimum and maximum vacuum noise
normalized variances to be: (2.0, 3.46) from
$\mathrm{OPA}_1$, (0.54, 5.16) from $\mathrm{OPA}_2$ and
finally (0.63, 2.54) from $\mathrm{OPA}_3$.  The phase-gates were
set to $\phi_1 = 90^{\circ}, \phi_2 = 41^{\circ}$ and $\phi_3 = 140^{\circ}$,
respectively. For further details, see Ref.\ \cite{EPAPS}.

The first OPA produces a classically squeezed (thermal)
state we refer to as \emph{hot squeezing}.  It manifests a non-uniform
stationary noise distribution amongst its two quadratures without having the
smallest quadrature fall below the vacuum noise level. Hot squeezing is
generated when, for example, two amplitude squeezed modes of different squeezing
factors are overlapped on a 50/50 beam splitter with a relative phase of $90^{\circ}$, thereby producing a two-mode squeezed state, but then one of
the output modes is discarded.
Without the presence of hot squeezing, we could not find any possibility to produce continuous-variable bound entanglement (between four modes), using the methods described in 
Ref.\ \cite{EPAPS}. Indeed, the passive optics following the sources can no longer
alter the eigenvalues of $\sigma\gamma$, which also define the degrees of squeezing
and mixedness of the state. Hot squeezing therefore appears to give rise to a 
necessary ingredient of quantum and classical correlations in order to create robust
bound entangled states.
We demonstrate that the same state can also be prepared {in a purely classical way} by
applying a local random displacement on the phase quadrature of a vacuum mode while
parametrically amplifying the state's amplitude quadrature.  The stationary
random phase modulation is produced by using an EOM (electro-optical modulator)
driven with the output 
from a homodyne detector
measuring shot noise.  The amplitude modulation is generated by operating
$\mathrm{OPA}_1$ in \fref{exp_setup} in amplification mode, effectively
anti-squeezing the amplitude quadrature and deamplifying the thermal noise phase
quadrature of the input state. In principle the random amplitude noise of the
first input mode can also be provided by a second homodyne detector and an
amplitude modulator, thereby replacing the parametric $\mathrm{OPA}_1$ device. We note that pseudo-random numbers {could be} insufficient in this scheme
since they could introduce artificial correlations and a non-stationary noise into the final state.

In order to hit the tiny regions in parameter space where bound entanglement does exist we introduce to our setup a new technique for precisely controlling
phase-gates at arbitrary angles.  This method relies on an optical single-sideband scheme (see 
{supplementary material}) that can be used to
arbitrarily and independently set the working point of both a phase-gate network
and multiple homodyne detectors.  This scheme reduces setting the relative phase
between interfering modes to selecting the electronic demodulation phase used in
the control loop.  A portion of the light leaving the phase-gates, PG1-3 in
\fref{exp_setup}, is redirected to control photodetectors.  We are able to
derive a strong error-signal by tapping only \unit[1]{$\mu$W} of power
corresponding to no more than 1\% of the signal mode's optical power.  For
applications where delicate quantum states must remain free from losses our
method provides a means by which they can still be used for controlled
interference without significant vacuum contribution due to loss.

\begin{figure}
\includegraphics{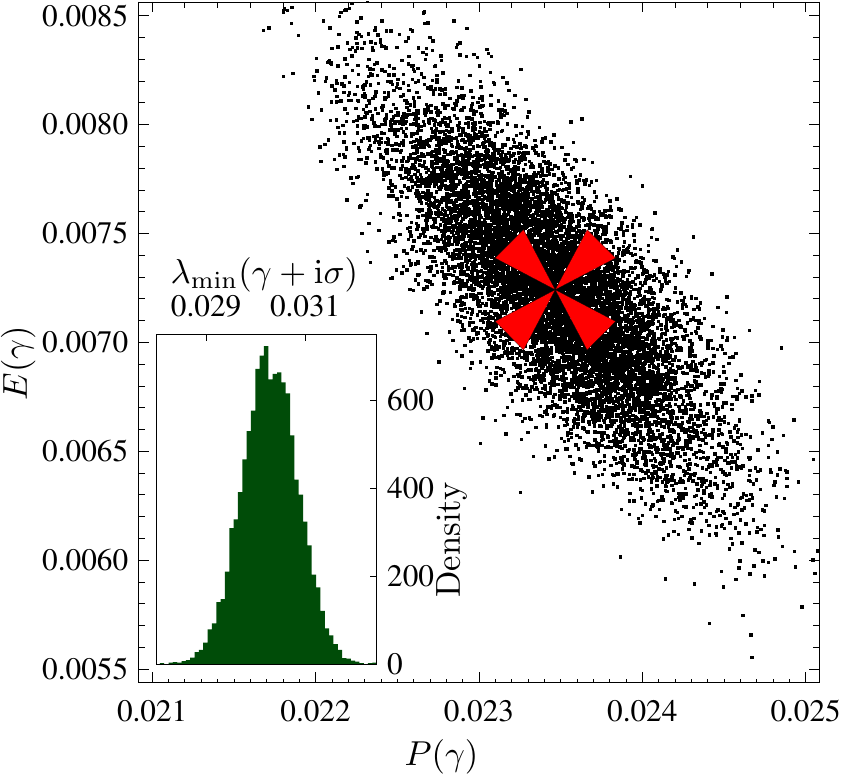}
\caption{Experimental results: The state measured after $4$ million sets
of  raw quadrature data points yields the entanglement $E$ and non-distillability
$P$ indicated by the red cross. Other $10^4$ points are obtained by
bootstrapping the original $4$ million data points and show that we are
$16\sigma$ away from separability and $46\sigma$ away from
distillability. In the inset we depict the minimum
eigenvalue of $\gamma+ \rmi\sigma$ of each of the $10^4$ bootstrapped
correlation matrices, showing that they are significantly far away from
the boundary of covariance matrices allowed by the uncertainty principle.}
\label{figure:experimentalResults}
\end{figure} 

The four balanced homodyne detectors are used for the full tomographic reconstruction of
the covariance matrix. The results of the reconstruction are used to evaluate
two characteristics of the state; namely, its entanglement $E$
(2) and its PPTness $P$ \eref{eq:P_def}.
In order to build the statistics of
these characteristics we first continuously recorded $4$ million data
points from the amplitude and phase quadratures of each mode. {Using the
bootstrapping method, we then randomly sampled from the total $4$ million 
points, with uniform distribution, points that were different}, and 
produced a series of covariance matrices from which the
entanglement, PPT and physical properties were {calculated}.  Our results are
represented in \fref{figure:experimentalResults} by the black points.  The 
cross corresponds to the average state inferred from the total data set. The
abscissa of \fref{figure:experimentalResults} is the PPTness and the ordinate
the entanglement.  By projecting the scatter plot onto the respective axes we
calculate a significance of $46\sigma$ away from being distillable, i.e.,
$P(\gamma) < 0$ and $16\sigma$ away from being separable, i.e., {$E(\gamma) \leq 0$}.
To demonstrate that the generated state is not close to the boundary
of state space (and to confirm its physicality), the smallest eigenvalue of 
$\gamma+ i\sigma$ is also depicted: This is 
shown in the inset as a histogram.
The fact that it is more than {$50\sigma$} away from being unphysical {can be seen
as} an indication of the fact that our setup was stable over the entire measurement
time and that our measured data exhibited little statistical uncertainty.

The fact that the involved states are Gaussian to a very high statistical significance
is carefully checked by statistical methods (see
supplementary material \cite{EPAPS}): Indeed, we have not only estimated the second
moments from the time series data from homodyning measurements, but in fact all moments.
Based on these data, we have computed Q-Q-plots for measured distributions against
perfectly Gaussian ones, showing a remarkable coincidence and confirming the precisely
Gaussian character of the state. What is more, from a bound of the mean energy of the 
state and of its Gaussian character, one can derive rigorous bounds to the distillable
entanglement of the state, confirming an at most negligible amount of distillable entanglement.
Again, for details, see the supplementary material \cite{EPAPS}.

Our results present the first unconditional preparation of bound entangled states of a physical system characterized by (continuous) position/momentum-like variables. With respect to systems composed of light, we demonstrate the first unconditional preparation of bound entanglement, and achieve an unprecedented significance of its features. Independent of any postselection, our platforms allows for the distribution of the entangled states. As other states of light our bound entangled states can be distributed to remote parties, which might be kilometers apart using optical fibers \cite{Fasersagnac10}. The decoherence on bound entangled states due to photon loss and phase noise \cite{Franzen07}  and the ineffectiveness of distillation schemes \cite{Hage08} can be tested, as well as the applicability of thermodynamical pictures of entanglement be studied experimentally.

Our results clearly exemplify the potential of the continuous variable platform for the precise engineering of complex multi-mode
states of light. We {underline} that using this platform the state preparation efficiency does not depend on the
number of entangled modes. That is to say, detecting, for example, one squeezed
mode with one homodyne detector has exactly the same efficiency as detecting $N$
squeezed states with $N$ homodyne detectors simultaneously.  Furthermore, we
estimate our total quantum detection efficient to be between $90$-$95$\% being
already considered in the preparation of bound entanglement.  
Alternatively, this loss could be mapped directly onto the measured state by inclusion of neutral density filters, and verification with perfect detectors would reveal the same statistics as depicted in \fref{figure:experimentalResults}. 

We believe that the precise and unconditional preparation of (bi-partite) bound entangled states of light demonstrated uplifts the theoretical and experimental research on the link between entanglement theory and  statistical physics. From a more general and also technological perspective, the high efficiency and the high degree of control in multimode quantum state preparation achieved certainly promotes the application of the unconditional continuous variable platform for the preparation of quantum states of light for fundamental research as well as quantum metrology.\\

This work has been supported by the EU (QESSENCE, MINOS, COMPAS),
the EURYI, the grant UNAM-PAPIIT IN117310 and by the Centre for Quantum
Engineering and Space-Time Research, QUEST. We acknowledge discussions
with P.\ Hyllus and M.\ Lewenstein at an early stage of this project.

\subsection*{Appendix A: Heisenberg uncertainty and entanglement criteria}

Explicitly, for $n$ modes the {\it symplectic matrix} $\sigma$ reads as
\begin{equation}
\sigma=\bigoplus_{j=1}^n\left(\begin{array}{cc}0 & 1 \\ -1 & 0\end{array}\right).
\end{equation}
The {\it Heisenberg uncertainty relation}, expressed in terms of the covariance matrix
\cite{simon1987}, is given by
\begin{equation}\label{eq:bonafide}
		 \gamma + \rmi  \sigma\geq 0.
\end{equation}
Such operator valued inequalities $A\geq B$ for Hermitian $A$ and $B$ always
refer to operator ordering, meaning that the real eigenvalues of $A-B$ are non-negative.
The entanglement measure $E$ for covariance matrices defined in the main text
indeed indicates entanglement in states \cite{EMeasure}, and for 
two modes this is essentially 
nothing but the familiar {\it negativity} 
\cite{Negativity1}.

In the discussion of the main text we show that the spectrum of $\gamma+
\rmi\sigma$ is bounded from below by $\varepsilon>0$, hence manifesting the
Heisenberg uncertainty principle. It is worth mentioning that this also means
that the smallest {\it symplectic eigenvalue} $s_1(\gamma)$ of $\gamma$ is
bounded away from $1$. In the experiment, we also test whether the
reconstructed covariance matrix satisfies the Heisenberg uncertainty relation
as this is a test if the matrix corresponds to a physical state. Unphysical
states might occur if the error bars of the quantum state preparation or the
tomographic characterization are too large.

\subsection*{Identifying robust bound entangled states}

The relative volume of bound entangled states compared to all states is very
small under every reasonable measure, and any verification as pursued here
necessarily requires a careful analysis as to what parameter set is most
suitable. In this subsection, we report techniques that have been used to
identify regimes of robust bound entangled states.

Consider a general $4$ mode correlation matrix, expressed in bipartite normal form \cite{napoli}:
\begin{equation}
\gamma = 
\begin{pmatrix} 
\lambda_1 & 0 & 0 & 0 		 		 	 & \lambda_5 & 0 &\lambda_9   & \lambda_{10} \\  
0 & \lambda_1 & 0 & 0 		 		 	 & 0 & \lambda_6 & \lambda_{11} & \lambda_{12} \\  
0 & 0 & \lambda_2 & 0 		 		 	 & \lambda_{13} & \lambda_{14} & \lambda_7 & 0 \\
0 & 0 & 0 & \lambda_2 		 		 	 & \lambda_{15} & \lambda_{16} & 0 & \lambda_8 \\  
\lambda_5 & 0 & \lambda_{13} & \lambda_{15}		 & \lambda_3 & 0 & 0 & 0 \\  
0 & \lambda_6 & \lambda_{14} & \lambda_{16}  		 & 0 & \lambda_3 & 0 & 0 \\  
\lambda_{9} & \lambda_{11} & \lambda_7 & 0  		 & 0 & 0 & \lambda_4 & 0 \\  
\lambda_{10} & \lambda_{12} & 0 & \lambda_8 		 & 0 & 0 & 0 & \lambda_4 \\  
\end{pmatrix}.
\label{eq:correlation_canonical}
\end{equation}
The $16$ parameters can be seen as describing a manifold in 
$\mathbb R^{16}$. We sample uniformly 
the hypercube $[-1/2,1/2]^{\times 16}$ until we get a bound entangled state. 
The evaluation of the $P$ measure for a given 
covariance matrix $\gamma$ amounts to solving an 
eigenvalue problem, that of the degree of entanglement $E$ to solving a 
{\it semi-definite problem}. In practice, the latter problem can also be 
performed in the dual space of witnesses that are quadratic polynomials
in the quadratures, as explored in Ref.\ \cite{P}.
Once an instance is found we construct a random walk in order to 
improve the robustness, by (i) displacements in the direction of the axis 
by a small amount $\delta\lambda=0.01$  and (ii) 
rotations, by the same angle, in each of the $16\times 15/2$ two 
dimensional Cartesian planes; the new covariance matrix is
accepted if the new corresponding state has a larger degree of entanglement {\it and}
is more significantly a PPT state as measured by $E$ and $P$, respectively. 
The most suitable state found (after several hundred hours of computer time), 
as quantified by the biggest value of $\min
\{E(\gamma), P(\gamma)\}$, 
is characterized by an entanglement value of
$E(\gamma)=0.054$ and $P(\gamma)=0.132$, giving an idea of the limiting values
that one can achieve.

However, experimentally it is too expensive to engineer a {state with an arbitrary} correlation
matrix. We thus construct a circuit which, starting from a product of noisy 
Gaussian single mode states, can produce bound entangled states, but is 
simple enough to be producible in the lab with available technology. 
A (non-unique) example of such a circuit is plotted in Fig.\ 1. {The resulting
scheme is a result of a variation within the given parameterized family of circuits --
again using a random walk approach as described above, but now 
on the physically feasible set of covariance matrices by parametrizing each
of the optical components -- 
maximizing the statistical significance of being bound entangled by running semi-definite
problems in each step.}
%
Afterwards we filter the results allowing only those which require achievable 
values of squeezing at the input and which only require a single mode with hot
squeezing, as this is also a precious resource that, at the moment, can only be
input in a single mode. Within the resulting states we choose the most robust
according to the aforementioned criteria. 

\subsection*{Appendix B: Details of the experiment}

The three OPAs used to produce the underlying quadrature squeezing at sideband
frequency of \unit[6.4]{MHz} were constructed from a type I
non-critically phase-matched $\textrm{MgO:LiNbO}_3$ crystal inside a standing
wave resonator, similar to the design that previously has been used in
Ref.\ \cite{chelkowski07}.  They were pumped with approximately \unit[100]{mW}
of green light at \unit[532]{nm} each resulting in a classical gain of about $5$.
The length of the OPA cavity as well as the phase of the second harmonic pump
beam were controlled using radio-frequency modulation/demodulation techniques.

Balanced homodyne detection was performed on each of the four modes in order to
reconstruct the $8\times8$ covariance matrix.  The optical local oscillator was
filtered through a three mirror ring cavity operated in high finesse mode
resulting in a linewidth of \unit[55]{kHz}.  The detector difference currents
were electronically mixed with a \unit[6.4]{MHz} local oscillator and
low-pass filtered with a \unit[400]{kHz} bandwidth. {The dark noise
separation from shot noise was measured to be more than \unit[10]{dB} for each
detector.} The raw data was acquired using a \unit[14]{bit} National Instruments DAQ-card
and in total eight measurement settings including the shot noise measurement
were required in order to reconstruct the covariance matrix.  

The hot squeezed states were generated by randomly phase modulating the control
beam used to set the length of the OPA cavity at the squeezing sideband
frequency, \unit[6.4]{MHz}, and locking the OPA cavity in amplification.  This
produces phase squeezed states whose smallest quadrature can be controlled by
varying the strength of the random noise modulated on the control field and
whose amplitude quadrature is controlled by the degree of classical gain.

The single-sideband was generated by overlapping the output of a second laser
operating at around \unit[1064]{nm} with the bright output of OPA1.  The beams
were phase-locked at a beat frequency of \unit[15]{MHz} resulting in a field
that corresponds to both a phase and amplitude modulation.  The beat was
detected by directing approximately 1\% of the phase-gate outputs to
photodetectors placed behind the phase-gates as well as in each homodyne
detector.  The relative phase between the carriers at both the phase-gates and
the homodyne detectors could then be set to an arbitrary phase simply by
changing the demodulation phase of the electronic local oscillator.  We estimate
a phase sensitivity at each phase-gate to be approximately \unit[2]{deg}.  

\subsection{Appendix C: Discussion of the Gaussian character of the state}

The findings of our work show with a remarkable and unprecedented 
statistical significance that
the second moments are certified to be those of 
Gaussian bound entangled states. 
What is more, a perfectly Gaussian statistics is clearly expected due to the underlying physical mechanisms of our setup, i.e., parametric amplification of vacuum states, their superposition on beam splitters, time-independent linear losses and phase shifts, and finally balanced homodyne detection. In this subsection, however, we highlight 
the extent to which the states are indeed Gaussian states, and carefully
discuss some theoretical issues associated with the certification of bound entanglement
for infinite-dimensional quantum systems.

\begin{figure}[tbh]
\centerline{\includegraphics[width=0.5 \textwidth]{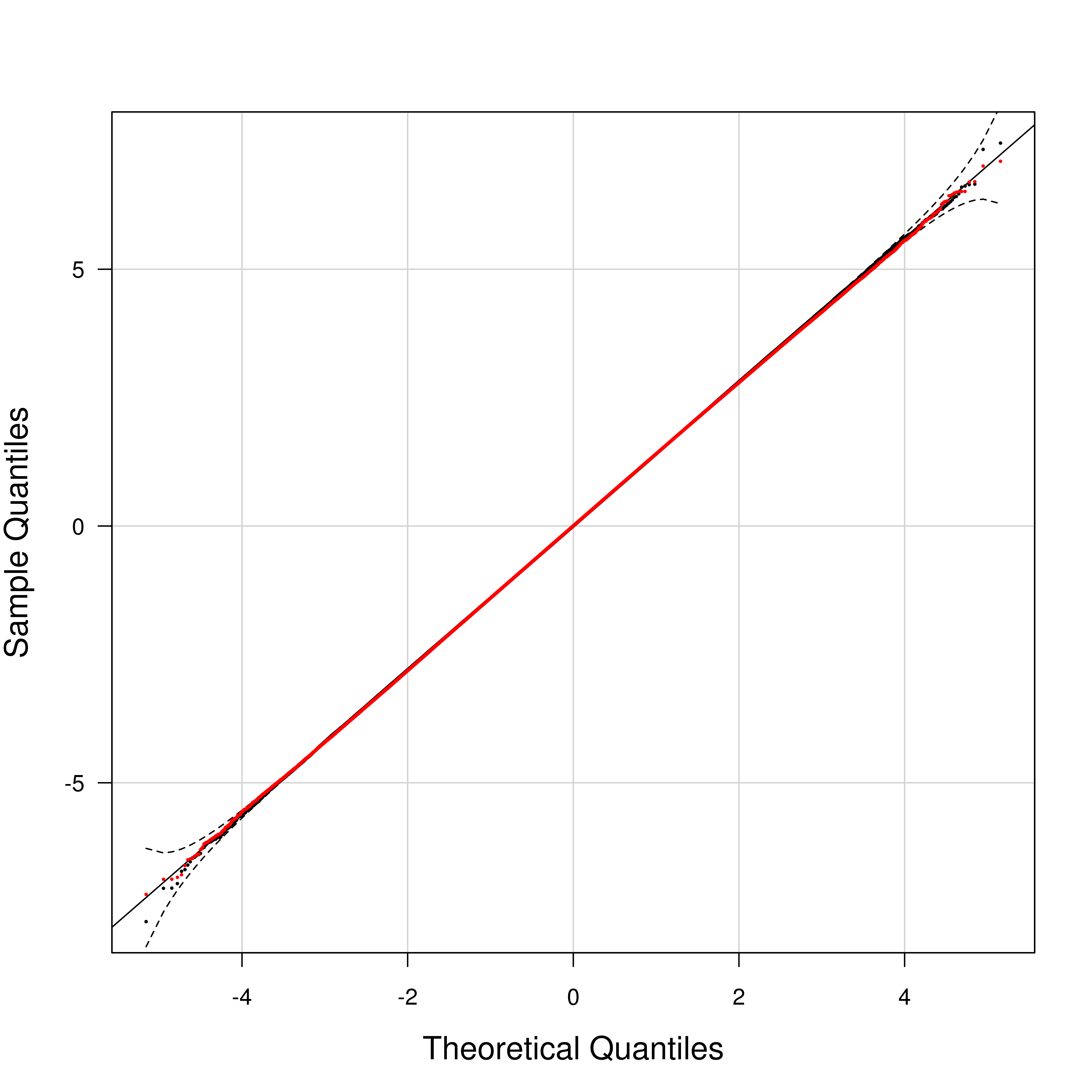}} 
\caption{Q-Q-plot of the sample quantiles of a measured 4 million point data set versus the 
theoretical one for a Gaussian distribution based on the measured second moments (red dots). For comparison we also give a Q-Q-plot by plotting quasi-random numbers from a perfect Gaussian distribution (black dots).  Such a Q-Q-plot depicts the $q$-th quantile of one distribution against the $q$-th quantile of the other. The diagrams show that the prepared 
states of our work are Gaussian to a very high level of significance.}
\label{f1}
\end{figure}

In the course of our work we have taken time series data from the homodyning measurements
and have not only estimated the second moments (leading to the covariance matrix specified
in the main text), but in fact all moments. For each moment, we have tested for the Gaussian character
of the state. Indeed, such
studies of Gaussianity are interesting in their own right.
More precisely, for our experiment, we have computed Q-Q-plots for a measured 
distribution against a Gaussian one, comparing these
two probability distributions by plotting their quantiles against each other.
Fig.\ \ref{f1} shows a representative sample Q-Q-plot of this kind. In each case
the Gaussian character is verified to a remarkable level of statistical significance. 
We also performed a $\chi^2$ test which also confirmed normality.
Hence, one can conclude that the prepared is indeed Gaussian to
a very high degree of accuracy.

Having said that, strictly speaking, one may argue that in infinite-dimensional
Hilbert spaces, the free-entangled states are (trace-norm) dense in state space
\cite{Dense,MyOldPaper}. Hence, in the vicinity of every bound entangled Gaussian
state one can find a free entangled state that is operationally indistinguishable. 
This is as such no surprise at all: In the same way, it is true that 
arbitrarily close to any separable state there is a free entangled state for continuous-variable
systems. However, in quantitative terms, the degree of free entanglement will indeed
be negligible \cite{BoundFootnote}. 
That is to say, the measured mean energy and the closeness to Gaussian states
readily give rise to rigorous bounds to the distillable entanglement of the unknown
state. Therefore, even in quantitative terms, one can falsify the proposition 
that a significant distillable entanglement is
found -- again strengthening the claim made in the main text.


\begin{thebibliography}{99}
\bibitem{BoundFootnote}	
	This can be seen as follows:
	Bounds of Gaussianity of a state as above imply that 
	$\|\rho-\rho_G\|_1\leq \varepsilon$ for a small $\varepsilon$, 
	where $\rho_G$ is the bound entangled Gaussian state 
	with 
	known covariance matrix and $\|.\|_1$ denotes the trace-norm. In turn, it has been shown
	in Ref.\ \cite{MyOldPaper} that the relative entropy of entanglement 
	$E_R$ with respect to separable states, 
	defined as the infimum $E_R(\rho)= \inf_{\omega} S(\rho\|\omega)$ over separable
	states $\omega$, is trace-norm continuous 
	for states with finite mean energy. What is more, $E_D(\rho)\leq E_R(\rho)$, so the 
	relative entropy of entanglement is always an upper bound to the distillable 
	entanglement $E_D$. This gives a bound to the distillable entanglement.
\end{thebibliography}
\end{document}